\documentclass[amsmath,amssymb,aps,reprint,prl,floatfix,superscriptaddress]{revtex4-2}
\usepackage{graphicx}
\usepackage{dcolumn}
\usepackage{bm}
\usepackage{amsmath}
\graphicspath{{figure/}}
\usepackage[nooneline,FIGTOPCAP]{subfigure}
\usepackage[colorlinks,linkcolor=red,anchorcolor=green,citecolor=blue]{hyperref}
\begin{document}

\preprint{}

\title{Skyrmion Dynamics in the Presence of Deformation}

\author{Zehan Chen}
\affiliation{Department of Physics, The Hong Kong University of Science and Technology, Hong Kong, China}

\author{Xichao Zhang}
\affiliation{Department of Electrical and Computer Engineering, Shinshu University, 4-17-1 Wakasato, Nagano 380-8553, Japan}

\author{Yan Zhou}
\affiliation{School of Science and Engineering, The Chinese University of Hong Kong, Shenzhen, Guangdong 518172, China}

\author{Qiming Shao}
\email[]{eeqshao@ust.hk}
\affiliation{Department of Physics, The Hong Kong University of Science and Technology, Hong Kong, China}
\affiliation{Department of Electronic and Computer Engineering, The Hong Kong University of Science and Technology, Hong Kong, China}
\affiliation{Guangdong-Hong Kong-Macao Joint Laboratory for Intelligent Micro-Nano Optoelectronic Technology, The Hong Kong University of Science and Technology, Hong Kong, China}


\date{\today}

\begin{abstract}
Magnetic skyrmions are topological spin textures promising for future high-density and non-volatile memory. It is crucial to understand the current-driven skyrmion dynamics in the presence of deformation, of which an analytical model, however, remains elusive. Here we extend Thiele's model by considering both the radial and tangential forces. Our model attributes the skyrmion deformation to the current-induced rotational symmetry breaking, which includes magnetization canting and domain wall width variation. Our predictions of skyrmion radius and nonlinear dynamics are consistent with micromagnetic simulation results. Besides, we show that by applying an in-plane magnetic field, the deformation of a skyrmion can be suppressed, and even the compression of a skyrmion can be achieved. Our model provides a generic way to analyze the skyrmion deformation and may inspire applications based on nonlinear skyrmion dynamics.
\end{abstract}


\maketitle

\textit{Introduction.—}
Skyrmion is a topological soliton first proposed by Skyrme to solve the nonlinear sigma model \cite{MesonsBaryons:Skyrme:NuclearPhysics}.
Recently, magnetic skyrmions were generated and observed in B20 chiral magnets and ferromagnet/heavy metal bilayer systems with Dzyaloshinskii-Moriya interaction (DMI) \cite{SpontaneousSkyrmionGroundState:Robler:Nature,SkyrOnTrack:Fert:NatNano,TopoDynmSkyrmions:Nagaosa:NatureNano,SpinOrbitTorqueDrivenSkyrmionDynamicsRevealedByTimeResolvedXRayMicroscopy,BlowingMagneticSkyrmionBubbles,RealSpaceObservationOfATwoDimensionalSkyrmionCrystal,SkyrmionLatticeInAChiralMagnet}.
Due to its non-trivial topological charge, skyrmion is topologically protected and can be driven by spin-transfer torque (STT) at ultralow current densities \cite{TheoryOfIsolatedMagneticSkyrmions,STTinMnSi:Jonietz:Science,UniversalCurrentVelocityRelationSkyrMotion:Iwasaki:NatComm}, promising for information carrier in non-volatile memory and other spintronic devices \cite{fert2017magnetic,RoomTempSkyrShiftDevice:Yu:NanoLett,SkyrmionBasedArtificialSynapsesForNeuromorphicComputing,SkyrmionElectronics:Zhang:JPCM}.

Manipulation of skyrmions can be achieved by current injection in a ferromagnet/heavy metal bilayer \cite{DirectObservationOfTheSkyrmionHallEffect,SkyrmionHallEffectRevealedByDirectTimeResolvedXRayMicroscopy,CurrentDrivenSkyrDynmSkyrHallEff:Juge:PhysRevApplied,TheRoleOfTemperatureAndDriveCurrentInSkyrmionDynamics}.
The charge current in the heavy metal layer with strong spin-orbit coupling induces a spin current in out-of-plane direction,
exerting spin-orbit torque (SOT) on the ferromagnetic layer to drive the skyrmion. For high-speed skyrmion applications, high drive current densities are needed.
Experimental results have suggested that skyrmion deformation can occur during the large current-driven motion \cite{TheRoleOfTemperatureAndDriveCurrentInSkyrmionDynamics,CurrentDrivenSkyrDynmSkyrHallEff:Juge:PhysRevApplied,DeformMovSkyrMnSi:Okuyama:Comm.Phys}.
Theoretically, STT- or SOT-induced skyrmion deformation has been studied using micromagnetic simulations and semianalytical approach  \cite{STTDrivenMotionDeformInstabSkyrHighCurrent:Masell:PhysRevB,OrientDepSkyrVariousTopo:Weissenhofer:PhysRevB,CurrentDrivenSkyrMotBeyondLinRegime:Liu:PhysRevApplied}.
Semianalytical means that micromagnetic simulation is still needed in analytical modeling.

Although the Thiele's approach \cite{SteadyStateMotion:Thiele:PhysRevLett} is very successful in explaining the current-driven skyrmion motion at low drive currents \cite{NucleationStabMotionSkyr:Sampaio:NatNano,UniversalCurrentVelocityRelationSkyrMotion:Iwasaki:NatComm,StrategyDesignSkyrMemory:Tomasello:ScientificReports}, and it has been extended to include antiferromagnetic spin textures \cite{zhang2016magnetic,barker2016static}, magnon effect \cite{SkyrmionDynamicsAtFiniteTemperaturesBeyondThieleEquation}, etc.,
there is no pure analytical model describing the deformation and nonlinear dynamics of current-driven skyrmions.
It is highly desired to extend Thiele's approach to better understand the skyrmion dynamics in the presence of high current-induced deformation \cite{SkyrmionsGetPushedBeyondTheLimit}.
In this work, we establish an analytical model including current-induced radial and tangential forces and successfully predict skyrmion radius and critical current density without the need of micromagnetic simulation.
Exceeding this critical current density, the skyrmion breaks down.
Induced by tangential Thiele's force density, the magnetization canting contributes the most to the symmetry breaking.
We also find that the domain wall width depends on the skyrmion radius.
Unlike most studies that focus on the translational part integration of Thiele's equation leading to skyrmion motion velocity, we calculate the integration of the radial part of Thiele's equation and obtain skyrmion radius.
With the obtained skyrmion radius, we can accurately describe nonlinear skyrmion dynamics.
Moreover, we propose a method to suppress the deformation by applying an in-plane magnetic field.

\textit{Model and Methods.—}
The dynamics of a ferromagnetic system is governed by the Landau–Lifshitz–Gilbert (LLG) equation \cite{gilbert2004phenomenological} with a damping-like SOT term,
\begin{equation}
    \frac{\partial\bm{m}}{\partial t}=-\gamma_0\bm{m}\times\bm{H}^{\rm eff}+\alpha\bm{m}\times\frac{\partial\bm{m}}{\partial t}-\tau_{\rm ad}\bm{m}\times(\bm{m}\times\bm{p})\label{eq:LLG},
\end{equation}
where $\gamma_0$ is the gyromagnetic ratio, $\bm{H}^{\rm eff}$ is the effective field, $\alpha$ is the Gilbert damping constant, $\tau_{\rm ad}$ is the damping-like spin orbit torque and $\bm{p}$ is the polarization direction.
Assume that $\bm{p}$ is along $+y$ direction, i.e. $\bm{p}=\bm{\hat{\jmath}}$ and electron current along $+x$ direction, i.e. $\bm{j}_e=j_x\bm{\hat{\imath}}$, where $\bm{\hat{\imath}},\bm{\hat{\jmath}}$ and $\bm{\hat{k}}$ are unit vectors in Cartesian coordinates.

Using Thiele's approach \cite{SteadyStateMotion:Thiele:PhysRevLett,GenHallEffSkyr:Wang:PhysRevB}, we can substitute each term in the LLG equation with an equivalent field, and define corresponding force densities, $f_i^\alpha=-\mu_0M_S\left(\partial_i\bm{m}\cdot\bm{H}^\alpha\right)$, $i=x,y$, $\alpha$ runs $g,e,d,c$.
$\bm{H}^g,\bm{H}^e,\bm{H}^d,\bm{H}^c$ are the effective fields corresponding to the terms in Eq.(\ref{eq:LLG}), and the exact definitions are given in Eqs.(S5)-(S8) in supplemental material \cite{Supplemental}.

A skyrmion in steady-state motion must satisfy $\bm{f}^g+\bm{f}^e+\bm{f}^d+\bm{f}^c=\bm{0}$ everywhere,
where $\bm{f}^g$ is gyroscopic force density, $\bm{f}^e$ is effective-field force density, $\bm{f}^c$ is current-induced SOT force density, and $\bm{f}^d$ is dissipative force density. Moreover, these forces can be decomposed into radial force and tangential force, $\bm{f}=f_r\bm{\hat{r}}+f_\varphi\bm{\hat{\varphi}}$,
where $\bm{\hat{r}}$ is the radial unit vector and $\bm{\hat{\varphi}}$ is the tangential unit vector.

For simplicity, we define radial force, which is the integral of radial force density, as,
\begin{equation}
    F_r^\alpha=\int_{\Sigma}\bm{f}^\alpha\cdot\bm{\hat{r}}{\rm d}\bm{S}=\int_0^{2\pi}\int_0^\infty\bm{f}^\alpha\cdot\bm{\hat{r}}{\rm d}r{\rm d}\varphi\label{eq:def:RadForce},
\end{equation}
where $\Sigma$ is the whole plane and $\bm{\hat{r}}$ is the unit radial vector in the polar coordinates centered at the skyrmion center (the center of mass).
Since all the force densities balance each other, their radial force density integrals or radial forces must balance as well, that $F_r^g+F_r^e+F_r^d+F_r^c=0$.
$F_r^g$, $F_r^d$ and $F_r^c$ are related to current and skyrmion motion, which cause the deformation of skyrmion.
So we define the expanding force to be the sum of these three radial forces, $F_r^{\rm exp}=F_r^g+F_r^d+F_r^c\label{eq:def:ExpForce}$.
To the contrary, $F_r^e$ is the restoring force, and we let $F_r^{\rm res}=F_r^e$.
It will be shown later that both $F_r^{\rm exp}$ and $F_r^{\rm res}$ depend on skyrmion radius $R$ and domain wall width parameter $w$.
By solving
\begin{equation}
    F_r^{\rm exp}+F_r^{\rm res}=0\label{eq:Model},
\end{equation}
we are able to find $R$ and $w$. How to accurately solve Eq.(\ref{eq:Model}) without using micromagnetic simulation is the major contribution of this work.

\begin{figure}
    \centering
    \includegraphics[scale=0.205]{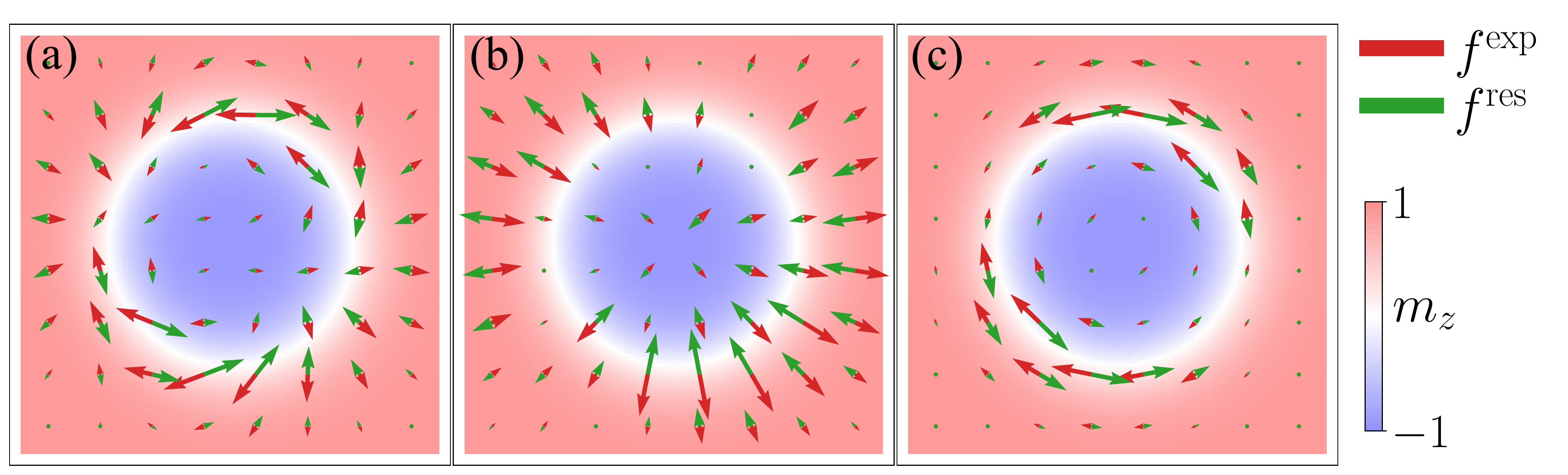}
    \caption{(a) The expanding force density $\bm{f}^{\rm exp}=\bm{f}^g+\bm{f}^c+\bm{f}^d$ and the restoring force density $\bm{f}^{\rm res}$ balancing each other in steady-state motion.
    (b) The radial parts and (c) the tangential parts of the expanding force density. The radial parts relate to skyrmion radius variation and the tangential parts relate to the magnetization canting.}
    \label{fig:forceDensity}
\end{figure}

The restoring force $F_r^{\rm res}=F_r^e$ is determined by effective field $\bm{H}^{\rm eff}$.
$\bm{H}^{\rm eff}$ contains four components, i.e., $\bm{H}^{\rm eff}=\bm{H}^{\rm exch}+\bm{H}^{\rm DMI}+\bm{H}^{\rm anis}+\bm{H}^{\rm ext}$,
which correspond to Heisenberg exchange interaction, DMI, effective anisotropy, and external magnetic field.
$\bm{H}^{\rm exch}=(2A/\mu_0M_S)\nabla^2\bm{m}$, $\bm{H}^{\rm anis}=(2K/\mu_0M_S)(\bm{m}\cdot\bm{\hat{k}})\bm{\hat{k}}$,
where $A$ is the exchange stiffness, $K$ is the effective anisotropy constant.
Effective anisotropy contains magnetocrystalline anisotropy and demagnetization field in ultrathin film with perpendicular magnetization \cite{SkyrmionConfinementUltrathinFilm:Rohart:PhysRevB}, that $K=K_u-\frac{1}{2}\mu_0M_S^2$, where $K_u$ is the uniaxial anisotropy constant.
For Néel-type skyrmion stabilized by interfacial DMI, $\bm{H}^{\rm DMI}=(-2D/\mu_0M_S)[(\nabla\bm{m})\bm{\hat{k}}-\nabla(\bm{m}\cdot\bm{\hat{k}})]$, and for Bloch-type skyrmion stabilized by bulk DMI, $\bm{H}^{\rm DMI}=(-2D/\mu_0M_S)\nabla\times\bm{m}$ \cite{StrategyDesignSkyrMemory:Tomasello:ScientificReports,SkyrmionConfinementUltrathinFilm:Rohart:PhysRevB}.

\textit{Rotational Symmetry Breaking of Skyrmion.—}
If a steady-state motion skyrmion takes a rotation-symmetric magnetization distribution similar to a ground state skyrmion, the expanding force $F_r^{\rm exp}$ will be zero.
SOT breaks the rotational symmetry of skyrmion, and induces a non-trivial $F_r^{\rm exp}$.
We will show that the breaking of rotational symmetry comes from two parts, i.e., magnetization canting and domain wall compression and expansion.

\begin{figure}[t]
    \centering
    \setlength{\abovecaptionskip}{0pt}
    \includegraphics[scale=0.28]{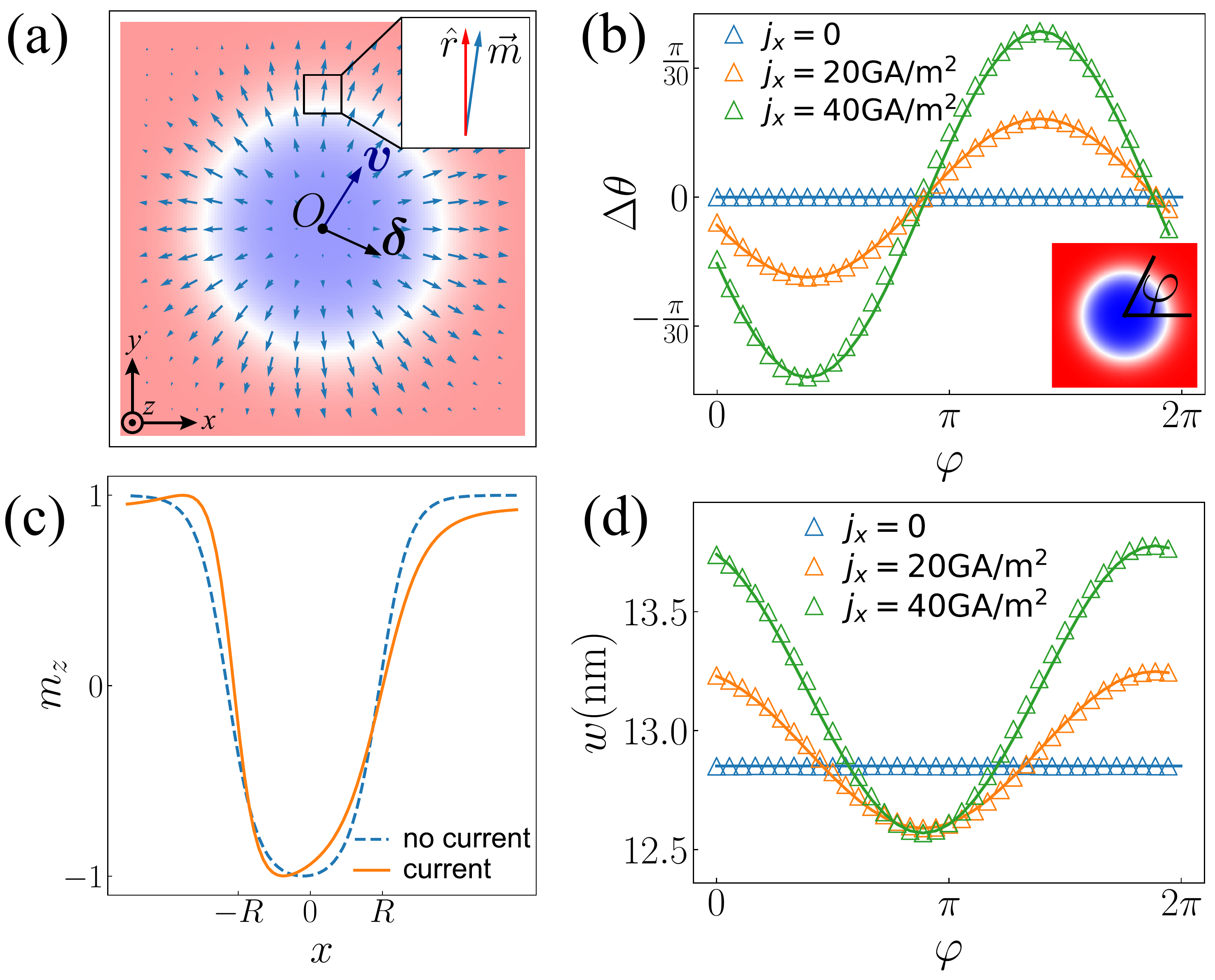}
    \caption{(a) Skyrmion canting configuration. In ground state, any magnetization should align with unit radial vector $\bm{\hat{r}}$, but canting occurs after applying current or in-plane field. A characteristic vector $\bm{\delta}$ is used to denote the amplitude and direction of canting, which is related to the amplitude and phase in (b).
    (b) Canting angle $\Delta\theta$ as a function of $\varphi$ under different current densities. The triangular dots are simulation results, and the solid lines are fitted sinusoidal curves. $\Delta\theta$ is defined as the angle between the radial unit vector $\bm{\hat{r}}$ and the in-plane $\bm{m}$ where $m_z=0$, as illustrated in (a).
    (c) Schematic of an intersection of moving and static skyrmion. $+x$ direction current injection will cause the left domain wall to compress and the right domain wall to expand.
    (d) $w$ as a function of $\varphi$ under different current densities. The triangular dots are simulation results, and the solid lines are fitted sinusoidal curves. $w$ is uniformly distributed only in ground state.}
    \label{fig:cantingAndDomainWall}
\end{figure}

Here, we discuss the Néel-type skyrmion case, where skyrmion domain wall profile is $m_z=\cos\Theta,m_r=\sin\Theta,\Theta(r)=2\arctan[\sinh(R/w)/\sinh(r/w)]$.
If no in-plane external field or current is applied, the direction of magnetization at each point of a ground state skyrmion will have no tangential component. 
However, a current will induce magnetization canting as shown in Fig.\ref{fig:cantingAndDomainWall}(a). The effect of an in-plane magnetic field is similar to that of a current \cite{STTDrivenMotionDeformInstabSkyrHighCurrent:Masell:PhysRevB}.

To quantitatively describe the canting, we use polar coordinates centered at skyrmion center and let $\boldsymbol{m}=m_r\bm{\hat{r}}+m_\varphi\bm{\hat{\varphi}}+m_z\bm{\hat{k}}$.
Note that the canting is equal to the tangential deviation of $\bm{m}$, i.e. $m_\varphi$. We show below that the canting is determined by tangential Thiele's force densities, which sum up to be zero as
\begin{equation}
    f_\varphi^c+f_\varphi^g+f_\varphi^d+f_\varphi^{\rm DMI}+f_\varphi^{\rm exch}+f_\varphi^{\rm anis}=0\label{eq:PhiPart}.
\end{equation}

Although Eq.(\ref{eq:PhiPart}) cannot be solved analytically, we can still estimate $m_\varphi$.
We ignore insignificant terms and obtain $m_\varphi=m_r\left(\bm{\delta}\cdot\bm{\hat{\varphi}}\right)$ \cite{Supplemental}, where
\begin{equation}
    \bm{\delta}=\bm{\delta}_{\rm current}=\frac{1}{2D}\frac{M_S}{\gamma}\left(\bm{v}\times\bm{\hat{k}}\right)\label{eq:def:delta}
\end{equation}
is the characteristic vector of canting shown in Fig.\ref{fig:cantingAndDomainWall}(a),
$\bm{v}=v_x\bm{\hat{\imath}}+v_y\bm{\hat{\jmath}}$ being the skyrmion velocity. It suggests that the canting is proportional to the skyrmion velocity. The canting vector $\bm{\delta}$, perpendicular to the velocity, characterizes the amplitude of the asymmetrical deformation of the skyrmion spin configuration.

With this estimation of $m_\varphi$, it can be shown that terms in the expanding force $F_r^{\rm exp}$ are \cite{Supplemental},
\begin{subequations}
\label{eq:det:Frexp}
\begin{eqnarray}
    F_r^c&&=\pi^2R\tau_{\rm ad}\frac{M_S}{\gamma}(\bm{\hat{\imath}}\cdot\bm{\delta})\label{subeq:det:Frc},\\
    F_r^g&&=2\pi\frac{M_S}{\gamma}(\bm{v}\times\bm{\hat{k}})\cdot\bm{\delta}\label{subeq:det:Frg},\\
    F_r^d&&=0.\label{eq:det:Frd}
\end{eqnarray}
\end{subequations}

The skyrmion velocity $\bm{v}$ can be computed by \cite{CurrentDrivenSkyrMotBeyondLinRegime:Liu:PhysRevApplied,GenHallEffSkyr:Wang:PhysRevB},
\begin{subequations}
\label{eq:det:velocity}
\begin{eqnarray}
    v_x&&=\tau_{\rm ad}\Lambda_{21}\frac{\alpha\Gamma_{11}}{(4\pi Q)^2+(\alpha\Gamma_{11})^2}\label{subeq:det:vx},\\
    v_y&&=\tau_{\rm ad}\Lambda_{21}\frac{4\pi Q}{(4\pi Q)^2+(\alpha\Gamma_{11})^2}\label{subeq:det:vy},\\
    Q=\pm 1&&,\quad\Lambda_{21}=\pi^2R,\quad \Gamma_{11}=2\pi\left(\frac{R}{w}+\frac{w}{R}\right).\label{subeq:factors}
\end{eqnarray}
\end{subequations}
Here $Q$ is the topological charge of skyrmion defined by $Q=\frac{1}{4\pi}\int_{\Sigma}\bm{m}\cdot\left(\frac{\partial\bm{m}}{\partial x}\times\frac{\partial\bm{m}}{\partial y}\right){\rm d}\bm{S}$ \cite{SkyrmionElectronics:Zhang:JPCM}, 
$\Lambda_{21}$ and $\Gamma_{11}$ are the driving factor and damping factor in Thiele's equation respectively.
Eq.(\ref{subeq:factors}) is derived from a rigid skyrmion as shown in Ref.\cite{CurrentDrivenSkyrMotBeyondLinRegime:Liu:PhysRevApplied}.

For current-driven skyrmion, $w$ is different from $w$ of ground state skyrmion. Also, $w$ along different azimuthal direction is not the same, i.e. $w=w(\varphi)$ is a function of polar angle $\varphi$, as shown in Figs.\ref{fig:cantingAndDomainWall}(c)(d). 
Without loss of validity, we can use average $w$, i.e., $w=\frac{1}{2\pi}\int_0^{2\pi}w(\varphi){\rm d}\varphi$.

The total Hamiltonian of a skyrmion includes exchange interaction, DMI, magnetocrystalline anisotropy and Zeeman energy due to external field. For ground state skyrmion without an external field, the skyrmion radius $R_0$ is given by $R_0 = \pi D \sqrt{A/(16AK^2-\pi^2 D^2 K)}$ \cite{TheorySkyrSize:Wang:CommPhys}.
The steady-state of skyrmion motion is a quasi-static state and the skyrmion should minimize its total energy under the constraint of Thiele's equation.
By minimizing the total energy, we have the relation between $R$ and $w$ for a moving skyrmion \cite{Supplemental},
\begin{equation}
    w=\sqrt{\frac{1}{\frac{1}{R^2}+\frac{K}{A}}}\label{eq:ConstraintW}.
\end{equation}

Unlike expectations in most papers that the domain wall is almost rigid \cite{CurrentDrivenSkyrMotBeyondLinRegime:Liu:PhysRevApplied}, we found that $w$ actually increases with $R$. This is also an improvement for normal ground state domain wall width parameter equation $w=\sqrt{A/K}$ \cite{SkyrmionConfinementUltrathinFilm:Rohart:PhysRevB}.

We then consider the effect of in-plane external field $\bm{B}^{\rm ext}=\mu_0\bm{H}^{\rm ext}$. It induces another canting vector $\bm{\delta}_{\rm ext}=wM_s\bm{B}_{\rm ext}/2D$ \cite{Supplemental}. Together with the current-induced canting, the net canting vector is $\bm{\delta}_{\rm net}=\bm{\delta}_{\rm current}+\bm{\delta}_{\rm ext}$.
If $-x$ direction in-plane field is applied, two canting vectors will partially cancel, and the expanding forces in Eq.(\ref{eq:det:Frexp}) will consequently be reduced and even be reversed.

In Eq.(\ref{eq:LLG}), we only consider damping-like SOT. Field-like SOT acts like an in-plane external field perpendicular to the current direction \cite{CurrentDrivenSkyrDynmSkyrHallEff:Juge:PhysRevApplied}.
Therefore, we can use an equivalent in-plane field to describe field-like SOT.

\begin{figure}[t]
    \centering
    \setlength{\abovecaptionskip}{0pt}
    \includegraphics[scale=0.21]{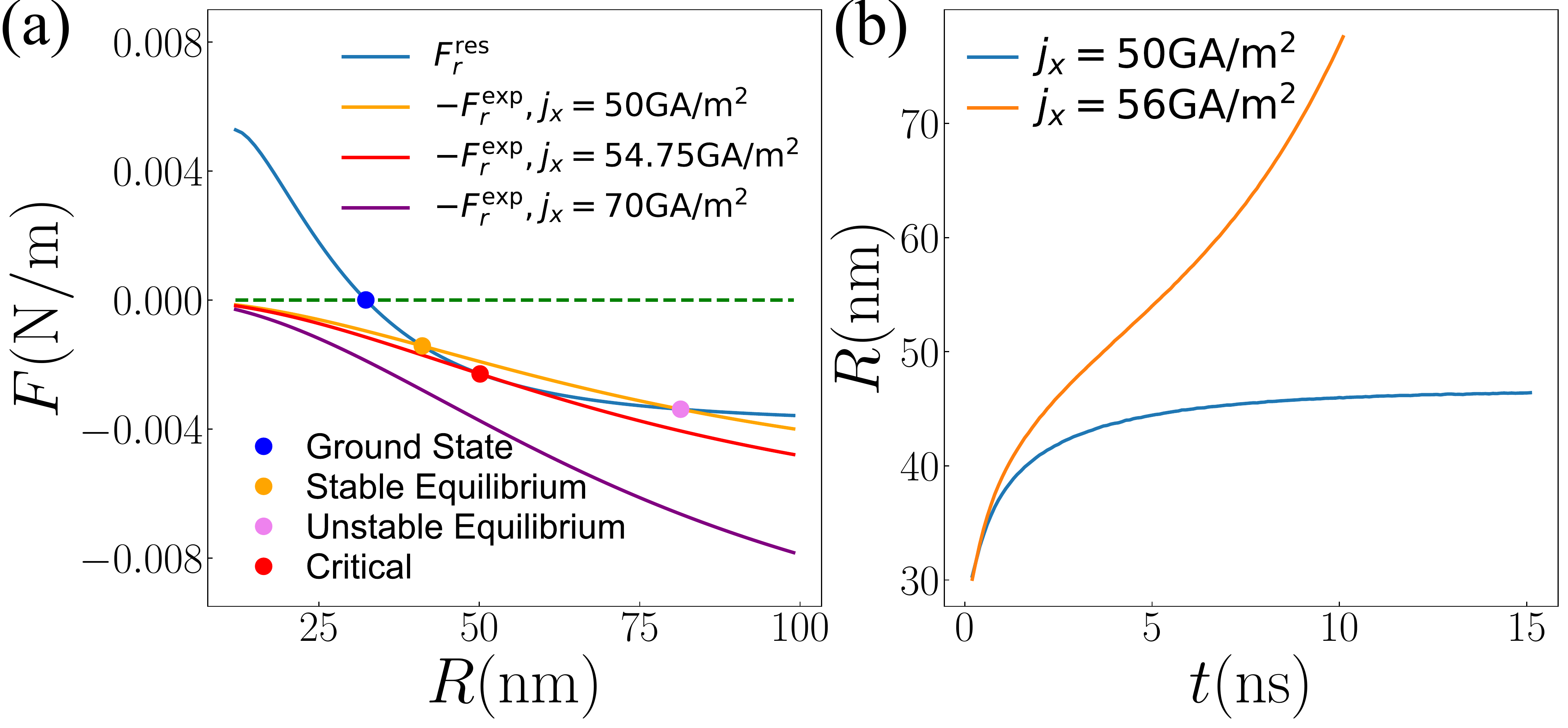}
    \caption{(a) Schematic of solving Eq.(\ref{eq:Model}). Solution exists for smaller $j_x$, but not for larger $j_x$.
    In the approximation, $F_r^{\rm res}$ and $F_r^{\rm exp}$ are treated as linear and quadratic respectively.
    (b) The skyrmion radii change with time in 15ns. Above critical current, skyrmion cannot have a saturation radius.}
    \label{fig:SkyrRadius}
\end{figure}

\begin{figure*}[t]
    \centering
    \setlength{\abovecaptionskip}{0pt}
    \includegraphics[scale=0.42]{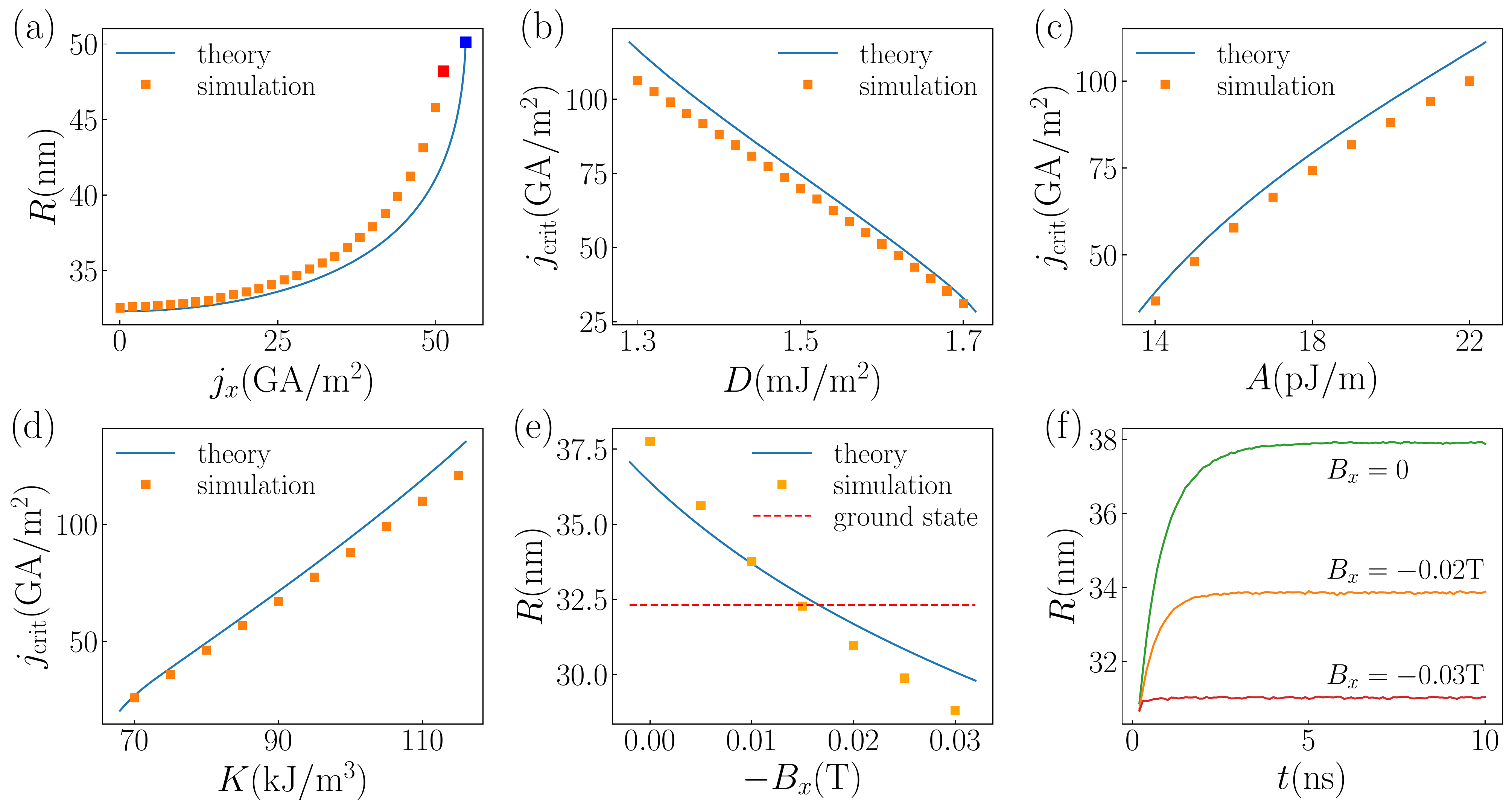}
    \caption{(a) Comparison of predicted radius and simulation results under different current densities.
    The red dot is the critical current and radius given by simulation. The blue dot is the critical point predicted by our model.
    The characteristics of the curve are captured by Eq.(\ref{eq:radiusApprox}).
    (b)-(d) Comparison between critical current density $J_{\rm crit}$ predicted by our model and the simulation results. $D=1.4\rm mJm^{-2}$ in (c) and (d).
    (e) The saturation radius as a function of a $-x$ direction in-plane external field $B_x$ under current density $j_x=40\rm GA/m^2$. The red dashed line represents the skyrmion radius with no current or in-plane field.  (f) The change of skyrmion radius with respect to time when different in-plane external fields and same current density are applied.
    }
    \label{fig:comparison}
\end{figure*}

To better understand the model, we can make approximations for $F_r^{\rm res}$ and $F_r^{\rm exp}$ and analytically solve Eq.(\ref{eq:Model}). The expanding force is approximately a quadratic function of both $R$ and $j_x$, $F_r^{\rm exp}=\beta j_x^2R^2$, where $\beta>0$. The restoring force $F_r^{\rm res}$ is a linear function of $R$, $F_r^{\rm res}=-\alpha_1(R/R_0-1)$, where $\alpha_1>0$. Both $\alpha_1$ and $\beta$ can be analytically determined \cite{Supplemental}, but their closed forms are too complicated to be written down. Approximately, $\alpha_1$ and $\beta D$ are constant \cite{Supplemental}. Eq.(\ref{eq:Model}) is therefore reduced to a quadratic equation, of which the physical solution is,
\begin{equation}
    R=\frac{\alpha_1-\sqrt{\alpha_1^2-4\alpha_1\beta j_x^2R_0^2}}{2\beta j_x^2R_0}\label{eq:radiusApprox}.
\end{equation}

The solution for $R$ only exists for $j_x$ smaller than some critical current density $j_{\rm crit}$, and $j_{\rm crit}=\sqrt{\alpha_1/4\beta}R_0^{-1}$.
The radius at $j_{\rm crit}$ is $R_{\rm crit}=2R_0$. 

In the presence of $x$-direction in-plane external field, the expanding force is attached with an additional term, $F_r^{\rm exp}=\beta j_x^2R^2+\lambda j_xB_xR$ \cite{Supplemental}. We can expand the effect of external field to the first order.
In such case, $j_{\rm crit}\approx\sqrt{\alpha_1/4\beta}R_0^{-1}-\lambda B_x/4\beta R_0$ and $R\approx R_c+\lambda j_xB_xR_0^2/\alpha_1$ \cite{Supplemental}, where $R_c$ is the skyrmion radius without external field, determined by Eq.(\ref{eq:radiusApprox}).
These approximations reveal that the critical current increases and the skyrmion radius can decrease if a sufficiently large $-x$ direction in-plane field is applied.

\textit{Verification Using Micromagnetic Simulation.—}
Since for fixed material parameters and $j_x$, $F_r^{\rm res}$ and $F_r^{\rm exp}$ only depend on $R,w$ and $\bm{v}$.
$w$ is determined by Eq.(\ref{eq:ConstraintW}), and $\bm{v}$ can also be calculated by parameters, $R$ and $w$ by Eq.(\ref{eq:det:velocity}).
As a result, numerically solving Eq.(\ref{eq:Model}) leads to $R$, the radius of the skyrmion in steady-state motion, called saturation radius.
However, when $j_x$ is lager than a threshold current density, skyrmion has no steady-state motion, as shown in Fig.\ref{fig:SkyrRadius}(a). We can find the critical current density by solving Eq.(\ref{eq:Model}). In Fig.\ref{fig:SkyrRadius}(b), we show two special cases, where one case with a drive current smaller than $j_{\rm crit}$ has a saturation radius and the other one with a drive current larger than $j_{\rm crit}$ does not. 

We use the Mumax3 software to perform micromagnetic simulations \cite{Mumax3::AIP}. The default parameters are $A=20{\rm pJm^{-1}}, K=0.1{\rm MJm^{-3}}, D=1.6{\rm mJm^{-2}}, M_S=1.42{\rm MAm^{-1}},\alpha=0.39$, thickness $d=0.9\rm nm$  \cite{CurrentDrivenSkyrDynmSkyrHallEff:Juge:PhysRevApplied,CurrentDrivenSkyrMotBeyondLinRegime:Liu:PhysRevApplied}.
We apply and adjust the $x$-direction current density, $j_x$, and record skyrmion saturation radius. Also, the existence of saturation radius is determined by testing whether the skyrmion radius diverges with time, as illustrated in Fig.\ref{fig:SkyrRadius}(b).
We compare the simulation results of current-dependent saturation radius with what our model predicts, and we achieve a good agreement, as shown in Fig.\ref{fig:comparison}(a).

We then verify our prediction for the critical current density $j_{\rm crit}$. $j_{\rm crit}$ is determined by the maximum $j_x$ such that solution still exists for Eq.(\ref{eq:Model}).
For each parameter setting, we use binary search to find the critical current using micromagnetic simulation.
We test for $D$, $A$ and $K$ respectively.
The results are shown in Figs.\ref{fig:comparison}(b)-(d) and are consistent with the predictions given by the analytical model. We can further understand the result from the approximation model, Eq.(\ref{eq:radiusApprox}).  $j_{\rm crit}=\sqrt{\alpha_1/4\beta}R_0^{-1}=\sqrt{\alpha_1/4\beta D}\left(\pi\sqrt{AD/\left(16AK^2-\pi^2D^2K\right)}\right)^{-1}$, where $\sqrt{\alpha_1/4\beta D}$ is constant \cite{Supplemental}. The dependence of $j_{\rm crit}$ on $D$, $A$ and $K$ can be easily understood using the approximation model now.

As predicted by the analytical model and the approximation model, applying an $-x$ direction in-plane field can suppress the deformation. Our predicted saturation radii and the micromagnetic simulation results in the presence of the external field are shown in Figs.\ref{fig:comparison}(e)(f), which show a quantitative agreement. Our approximation model provides the insight that in-plane external field can cancel the magnetization canting caused by current, suppress skyrmion deformation, and even cause skyrmion to contract under current injection. 

\begin{figure}[t]
    \centering
    \setlength{\abovecaptionskip}{0pt}
    \includegraphics[scale=0.24]{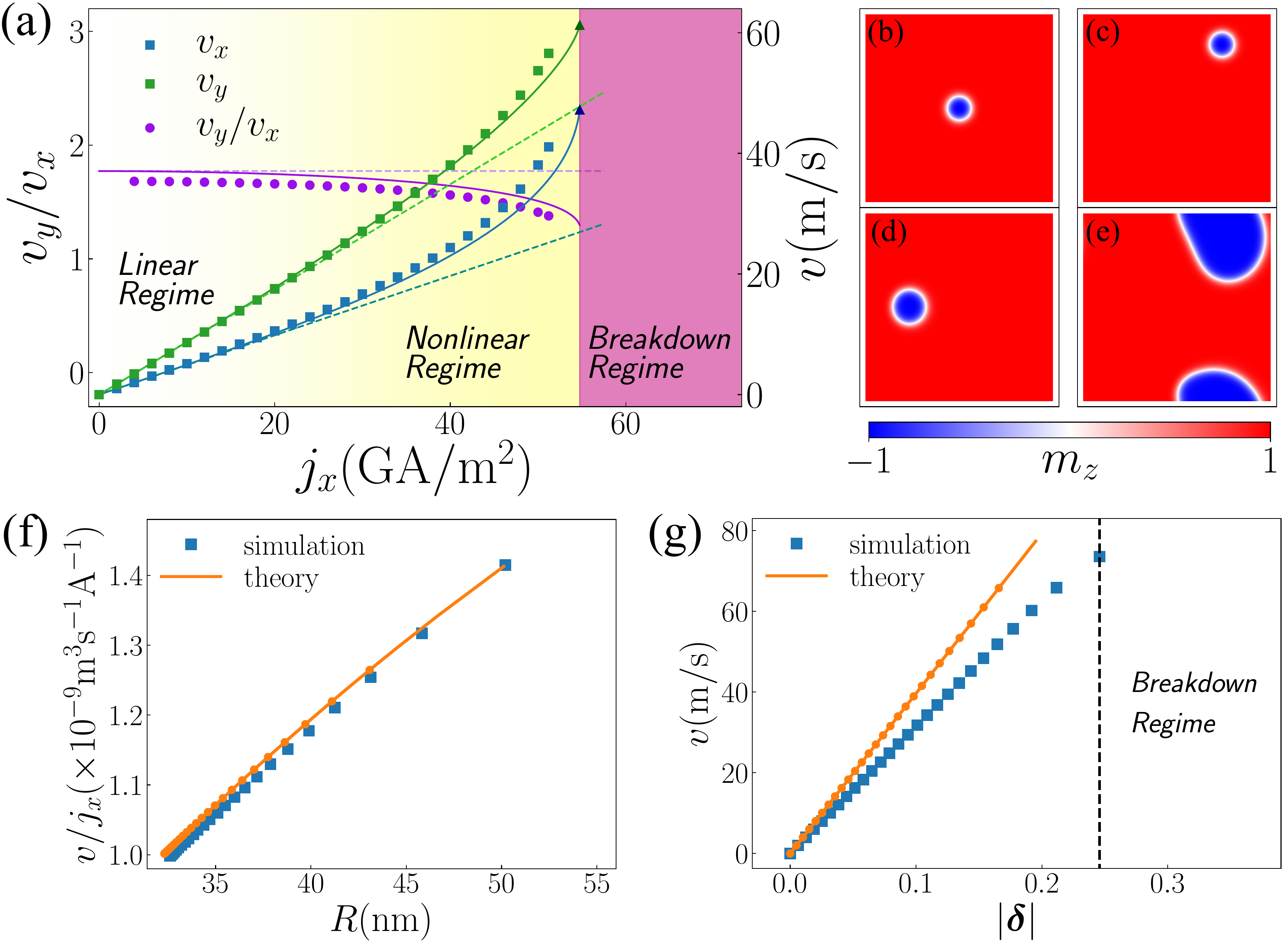}
    \caption{(a) Velocities ($v_x$ and $v_y$) and skyrmion Hall angle as functions of current density. Square dots are simulation results. Solid curves are prediction of our model. Dashed lines are normal estimations of skyrmion velocity and Hall angle assuming no deformation. The nonlinear response to a large current is well predicted by our model.
    (b) Ground state skyrmion, (c) skyrmion in linear regime, (d) skyrmion in nonlinear regime and (e) skyrmion in breakdown regime. Periodic boundary condition is used here. Skyrmion size continues to increase and the skyrmion eventually collapses. See supplemental movies S1-S3 for (c)-(e) \cite{Supplemental}.
    (f) Skyrmion mobility (defined as the skyrmion velocity divided by the current density) as a function of skyrmion radius. The mobility increases as the radius increases.
    (g) Skyrmion velocity as a function of the length of the canting vector $\bm{\delta}$. The circles on the theoretical curves in (f) and (g) have the same current density with those on the simulation curves.
    }
    \label{fig:velocity}
\end{figure}

At last, we describe the nonlinear velocity of skyrmion in the presence of deformation, which goes beyond Thiele's equation in the linear regime. The change in skyrmion radius induces nonlinear response of skyrmion velocity by Eq.(\ref{eq:det:velocity}), as shown in Fig.\ref{fig:velocity}(a). Our analytical model provides a much better predictions about velocity and skyrmion Hall angle than the linear model, where only translational parts of forces are considered. We also show snapshots of skyrmion dynamics at different regimes in Figs.\ref{fig:velocity}(b)-(e). We can see that our predicted velocity and skyrmion Hall angle are consistent with the micromagnetic simulation.
From the simulation results in Fig.\ref{fig:velocity}(f), we find that the mobility of the skyrmion, defined as $v/j_x$, increases as the skyrmion radius increases. This is consistent with our analytical model and Eqs.(\ref{eq:det:velocity}). To investigate how rotational symmetry breaking affects the skyrmion dynamics, we show the relation between the skyrmion velocity and the magnetization canting, which is the asymmetric deformation of the skyrmion spin configuration in Fig.\ref{fig:velocity}(g). The velocity increases quasi-linearly with the magnetization canting, which is consistent with the analytical model and Eq.(\ref{eq:def:delta}). The quantitative discrepancy between the simulated and the theoretical canting is expected to resolve by further improving the assumptions from Eq.(\ref{eq:PhiPart}) to Eq.(\ref{eq:def:delta}). 

In supplemental material \cite{Supplemental}, an empirical correction of the model is proposed, leading to a significant improvement in the quantitative agreement. Besides, we generalize our model to cases that involve out-of-plane external field and deformation dynamics of Bloch-type skyrmions, and show our model works very well for these cases.

In conclusion, our analytical and nonlinear model accurately describes the deformation and velocity of SOT-driven skyrmion. It captures the essential physics of skyrmion deformation, that is the rotational symmetry breaking of SOT-driven skyrmion. From this foothold we are able to suppress skyrmion deformation by applying in-plane field, which also has a sizeable effect on skyrmion symmetry. Our results suggest that the spatial symmetry breaking in topological spin textures may significantly affect their dynamics, which can be potentially utilized to build nonlinear functional devices \cite{ChaosAFMBimeron:Shen:PhysRevLett}.
Moreover, we anticipate that our model can be extended to describe nonlinear dynamics of other topological spin textures such as antiferromagnetic skyrmions and bimeron.

We appreciate fruitful discussions with Motohiko Ezawa and Se Kwon Kim. Z.C. acknowledges the support from Zheyu Ren and Cheuk Pan Fong. The authors at HKUST acknowledge funding support from the Shenzhen-Hong Kong-Macau Science and Technology Program (Category C, Grant No. SGDX2020110309460000), Research Grant Council—Early Career Scheme (Grant No. 26200520), and the Research Fund of Guangdong-Hong Kong-Macao Joint Laboratory for Intelligent
Micro-Nano Optoelectronic Technology (Grant No. 2020B1212030010). Y.Z. acknowledges the support by Guangdong Special Support Project (2019BT02X030), Shenzhen Fundamental Research Fund (Grant No. JCYJ20210324120213037), Shenzhen Peacock Group Plan (KQTD20180413181702403), Pearl River Recruitment Program of Talents (2017GC010293) and National Natural Science Foundation of China (11974298, 61961136006).X.Z. was an International Research Fellow of the Japan Society for the Promotion of Science (JSPS). X.Z. was supported by JSPS KAKENHI (Grant No. JP20F20363).

\bibliography{ref.bib}


\end{document}